\documentclass[reprint,amsmath,amssymb,aps,]{revtex4-2}

\usepackage{graphicx}
\usepackage{color}
\usepackage{dcolumn}
\usepackage{mathtools}
\usepackage{makecell,tabularx}
\usepackage{bm}

\begin{document}

\preprint{APS/123-QED}

\title{Autonomous systems and attractor behaviors in non-metricity gravity: stability analysis and cosmic acceleration}

\author{Pooja Vishwakarma}
 \email{poojavishwakarma041993@gmail.com}
 \affiliation{Department of Mathematics, School of Advanced Sciences, VIT-AP University,
Amaravati 522237, India}
\author{Parth Shah}%
 \email{parthshah2908@gmail.com}
\affiliation{%
SVKM's Narsee Monjee College of Commerce and Economics, Mumbai 400056, India 
}%
\author{Kazuharu Bamba}
 \email{bamba@sss.fukushima-u.ac.jp}
\affiliation{Faculty of Symbiotic Systems Science, Fukushima University, Fukushima 960-1296, Japan}

\begin{abstract}
The cosmological dynamics are rigorously investigated through the systematic application of autonomous system analysis to the gravitational field equations in non-metricity gravity. The systematic procedure to analyze the late-time cosmic acceleration in higher-order non-metricity gravity is demonstrated by exploring non-hyperbolic critical points with the center manifold theory. The stability properties of these critical points are also evaluated based on the analysis of eigenvalues and phase portraits. It is explicitly shown that the stable node can be realized. The critical points of each model are individually analyzed, and their corresponding cosmological implications are derived. The stability properties of these critical points are evaluated based on the analysis of eigenvalues and phase portraits, revealing that each model includes at least one stable node. Furthermore, the evolution plots of the cosmological parameters confirm the models' capacity to exhibit accelerated expansion.
\end{abstract}

\maketitle

\section{Introduction}
The quest to understand the fundamental nature of our universe has driven cosmologists to explore a variety of theoretical frameworks. A significant challenge that theoretical physicists have encountered is explaining the universe's late-time accelerated expansion. Numerous observational indicators, including Type Ia supernovae (SNIa) \cite{SupernovaCosmologyProject:1998vns, SupernovaSearchTeam:1998fmf, SupernovaSearchTeam:2004lze}, Large Scale Structure (LSS), the Cosmic Microwave Background (CMB), and Baryonic Acoustic Oscillations (BAO) \cite{Planck:2018vyg, Koivisto:2005mm, Daniel:2008et, SDSS:2005xqv}, have contributed to the current consensus that the universe is undergoing an accelerated phase of expansion.

The most extensively accepted theory posits that the universe is currently dominated by dark energy. However, none of the existing dark energy models are entirely satisfactory.The $\Lambda$CDM model, defined by the equation of state $w_{\Lambda} = -1$, stands as the leading contender among dark energy models \cite{Copeland:2006wr, padmanabhan2007, Durrer2008, Bamba:2012cp, Kase:2018aps}. Albert Einstein's formulation of General Relativity (GR) marked a significant shift in the understanding of gravity, redefining it as the curvature of spacetime resulting from mass and energy. Despite its successes in explaining a wide range of phenomena, GR faces several challenges, particularly in explaining the accelerated expansion of the universe, the nature of dark energy, and the behavior of gravity at quantum scales. These limitations have prompted physicists to consider alternative theories of gravity \cite{Ratra:1987rm, Caldwell:1997ii, Caldwell:1999ew, Feng:2004ad, Bean:2008ga, Langlois:2018dxi}. Modifying the geometry of spacetime is another technique to explain the universe's current acceleration.

To achieve this, we modify the Einstein-Hilbert action of General Relativity (GR). With the exception of non-Lagrangian theories such as Modified Newtonian Dynamics (MOND), modified theories of gravity are often represented by an altered Lagrangian density, which includes new geometrodynamical elements in the Einstein-Hilbert action integral \cite{Koivisto:2018aip}. Numerous such modified theories currently exist. Among them, $ f(R) $, $ f(R,T)$, $ f(T) $, and $ f(Q) $ gravities are the most successful in terms of cosmological viability, among which $ f(Q) $ gravity stands out due to its innovative approach and promising results. The $ f(R) $ gravity theory, expresses gravity as an arbitrary function of the Ricci scalar curvature $ R $. This concept was first introduced by H.A. Buchdahl \cite{Buchdahl:1970ynr} and further explored in the seminal works of V. Faraoni \cite{Sotiriou:2008rp}. This theory is of particular interest because it may provide a geometric method for describing inflation \cite{Starobinsky:1980te, Brooker:2016oqa, Huang:2013hsb} and addressing the dark energy problem \cite{Nojiri:2017ncd, Capozziello:2011et, Nojiri:2010wj, Arai:2022ilw}. In the $f(R,T)$ theory \cite{Harko:2011kv}, the gravitational action is extended by incorporating additional terms that involve both the Ricci scalar $R$ and the energy-momentum tensor $T$, which characterizes the distribution of matter and energy throughout spacetime.

An alternative approach to describing gravitational interactions involves incorporating torsion and non-metricity into the theoretical framework. Gravities derived from these characteristics are referred to as GR analogues, with modified versions known as $f(T)$ and $f(Q)$ gravities \cite{Capozziello:2011et, Clifton:2011jh, Capozziello:2023vne, Joyce:2014kja}. These gravitational theories can be formulated using non-standard metric-affine connections, such as the Weitzenbock connection and metric incompatible connections, which differ from the Levi-Civita connection used in General Relativity (GR). These frameworks explore gravitational dynamics beyond the conventional GR paradigm by incorporating diverse geometric structures and connections. In the realm of General Relativity, the Levi-Civita connection is associated with curvature while maintaining zero torsion. In contrast, the teleparallelism framework employs the Weitzenbock connection, which is characterized by torsion and zero curvature. This distinction between the two connections highlights their primary geometric characteristics and elucidates their respective roles in describing gravitational interactions within their theoretical contexts \cite{Koussour:2022zgo}. The $f(T)$ gravity is recognized as the most comprehensible teleparallel equivalent of General Relativity (TEGR) \cite{Ferraro:2006jd, Cai:2015emx, Bahamonde:2021gfp, Duchaniya:2023aeu}. 

In teleparallel gravity, two prominent theoretical challenges are frequently discussed. Firstly, there is a lack of local Lorentz symmetry, implying that the theory does not remain invariant under local Lorentz transformations. This issue raises concerns about the physical equivalence of different reference frames in the theory. Secondly, the existence of a ghost mode has been identified, which may introduce non-physical degrees of freedom leading to instability in certain regimes of the theory. Despite these challenges, teleparallel gravity remains an attractive framework as it offers a different geometric perspective on gravity, replacing curvature with torsion as the fundamental geometric quantity. It is also useful to explore as it allows for potentially novel extensions and modifications to General Relativity, which may lead to resolving cosmological issues such as the accelerated expansion of the Universe. For further details, see: \cite{Ferraro:2006jd, Aldrovandi:2013wha, Ong:2013qja, Rodriguez-Benites:2024pce}.

Recently, a novel gravitational theory known as the symmetric teleparallel equivalent of General Relativity (STEGR) has been investigated. This theory employs the concept of non-metricity scalar $Q$ to define gravitational interactions, characterized by zero torsion and curvature \cite{BeltranJimenez:2017tkd, BeltranJimenez:2019tme}. For a comprehensive review on $f(Q)$ gravity, see \cite{Heisenberg:2023lru}. Both teleparallel and symmetric teleparallel gravities can be developed within intriguing geometric frameworks provided by torsion and non-metricity, respectively \cite{Mandal:2020lyq, Mandal:2020buf, Harko:2018gxr, Dimakis:2021gby, Koussour:2022ycn, Koussour:2022wbi, Narawade:2023tnn, Agrawal:2022ccq, Pati:2021zew, Vishwakarma:2023brw, BeltranJimenez:2017tkd, Heisenberg:2023wgk, DAmbrosio:2023asf, Hu:2023gui, Gomes:2023tur}. In this study, we focus on the case of modified symmetric teleparallel gravity. The non-metricity scalar $ Q $ in $f(Q)$ gravity is a measure of how much the geometry of spacetime deviates from being purely metric. Unlike $ f(R) $ gravity, which modifies the Ricci scalar $ R $, $ f(Q) $ gravity modifies the non-metricity scalar, providing a different perspective on the geometric properties of spacetime. This modification has significant implications for cosmology, offering new ways to model the universe's expansion and structure formation without relying on dark energy or other exotic components.

Dynamical systems analysis is a powerful mathematical tool that allows for the qualitative study of cosmological models. By transforming the complex field equations into an autonomous system of differential equations, one can analyze the stability and behavior of cosmological solutions \cite{Odintsov:2017tbc, Odintsov:2017icc, Hohmann:2017jao}. This approach helps in identifying critical points, understanding their nature, and exploring the evolutionary paths of the universe within the theoretical model \cite{Odintsov:2018uaw, Odintsov:2019ofr}. The application of dynamical systems theory to cosmology is particularly valuable for several reasons:\\
(i)\textbf{Stability Analysis:} Identifying and analyzing critical points in the dynamical system helps determine the stability of various cosmological solutions. Stable solutions are particularly important as they can represent realistic models of the universe's long-term behavior.\\
(ii) \textbf{Qualitative Behavior:}  Dynamical systems analysis allows for the exploration of qualitative behaviors of cosmological models, such as the presence of attractors, repellers, and saddle points. These features provide insights into the possible evolutionary trajectories of the universe.\\
(iii) \textbf{Comparative Analysis:} By comparing different forms of the function $f(Q)$, one can study how various modifications to the non-metricity scalar affect the cosmological dynamics. This comparative approach is crucial for understanding the strengths and limitations of different models within the $ f(Q) $ framework.

For non-hyperbolic points, linear stability fails. A critical point, known as a non-hyperbolic point, if it  has zero real part among its eigenvalues (a critical point is considered hyperbolic if none of its eigenvalues is zero). In these cases, the stability properties of the system must be studied using other techniques, such as Center Manifold Theory and Lyapunov Functions \cite{Vishwakarma:2023brw, Vishwakarma:2024qvw, Das:2019ixt}. In this work, we are using ``Center Manifold Theory" to study the stability properties of non-hyperbolic points.

This manuscript aims to provide a comprehensive dynamical systems analysis of cosmological models at background and  perturbation level within the $f(Q)$ gravity framework. By systematically investigating the phase space of these models, we seek to uncover the conditions under which they exhibit stable, accelerated expansion solutions. Additionally, we will explore the impact of different functional forms of $ f(Q) $ on the dynamical behavior of the universe. These studies can be utilised to support the findings of the observational analysis. The paper is organised as follows: In Sec. II, we presents the $f(Q)$ gravity field equations, from which the background and perturbed cosmological equations can be obtained. In Sec. III, we describe the formalism of the Center Manifold Theory. In Sec. IV, we investigate the phase space analysis of the three models governed by the form of the function $ f(Q) $. Finally, the results are summarised in Sec. V. 

\section{Symmetric Teleparallel Gravity}

In $f(Q)$ gravity, the gravitational interaction is described by a function of the non-metricity scalar $ Q $. This formulation falls within the broader category of metric-affine theories, where both the metric $ g_{\mu\nu} $ and the affine connection $ \Gamma^\lambda_{\mu\nu} $ are treated as independent variables \cite{BeltranJimenez:2017tkd, BeltranJimenez:2019tme, Khyllep:2021wjd, Bahamonde:2017ize}.

Non-metricity is defined as the covariant derivative of the metric tensor, given by:
\begin{center}
$ Q_{\lambda\mu\nu} = \nabla_\lambda g_{\mu\nu} $.
\end{center}
The non-metricity tensor can be decomposed into two independent traces:
\begin{center}
$Q_\mu = Q_{\mu\;\;\nu}^{\;\;\nu}, \quad \tilde{Q}^\mu = Q_{\nu}^{\;\;\mu\nu}$.
\end{center}
The non-metricity scalar $ Q $ is then constructed from these traces as follows:
\begin{center}
$ Q = -g^{\mu\nu} (L^\alpha_{\;\;\beta\mu} L^\beta_{\;\;\nu\alpha} - L^\alpha_{\;\;\beta\alpha} L^\beta_{\;\;\mu\nu}) $,
\end{center}
where $ L^\lambda_{\;\;\mu\nu} $ are the disformation coefficients, defined in terms of the connection and the metric.

The action for $ f(Q) $ gravity is given by:
\begin{equation}
S = \int d^4x \sqrt{-g} \left( \frac{1}{2} f(Q) + \mathcal{L}_{\mathrm{m}} \right),
\end{equation}
where $ \mathcal{L}_{\mathrm{m}} $ is the matter Lagrangian density.
To derive the field equations, we vary the action with respect to the metric tensor $ g_{\mu\nu} $ and the connection $ \Gamma^\lambda_{\mu\nu} $.
The variation of the action with respect to the metric yields:
\begin{equation}
\begin{split}
\delta_{g} S =  \int d^4x \sqrt{-g} &  \left[ \frac{1}{2} f_{Q} \delta_g Q + \frac{1}{2} f(Q) g_{\mu\nu} \delta g^{\mu\nu}\right. \\
&+\left. \delta_g (\sqrt{-g} \mathcal{L}_{\mathrm{m}}) \right].
\end{split}
\end{equation}
Here, $ f_{Q} $ denotes the derivative of $ f(Q) $ with respect to $ Q $.
Using the chain rule and integrating by parts, we obtain the metric field equations:
\begin{equation}
\begin{split}
\frac{1}{2} g_{\mu\nu} f(Q) & + \frac{1}{2} f_{Q} \left[ -2Q_{(\mu\nu)\alpha}^{\;\;\;\;\;\;\;\alpha} + 2\nabla_\alpha (Q^\alpha_{\;\;(\mu\nu)} \right. \\
&  - \left. Q_{(\mu\nu)}^{\;\;\;\;\;\alpha}) \right] = T_{\mu\nu},
\end{split}
\end{equation}
where $ T_{\mu\nu} $ is the energy-momentum tensor.

The variation with respect to the connection yields:
\begin{equation}
\delta_\Gamma S = \int d^4x \sqrt{-g} \left[ \frac{1}{2} f_{Q} \delta_\Gamma Q \right].
\end{equation}
This leads to the connection field equations, which, after some algebra, can be expressed as:
\begin{equation}
\begin{split}
\nabla_\lambda (\sqrt{-g} f_{Q} g^{\mu\nu}) & - \frac{1}{2} \sqrt{-g} f_{Q} \left( g^{\mu\nu} Q_{\lambda} \right. \\ & - \left. g^{\mu\alpha} Q_{\alpha\lambda}^{\;\;\;\;\nu} \right) = 0.
\end{split}
\end{equation}

In this work, we shall strictly follow a spatially flat Friedmann-Lemaître-Robertson-Walker (FLRW) spacetime. This model is considered the benchmark for representing the Universe on a huge scale, assuming that it is uniform and isotropic. The metric for this spacetime is defined by:
\begin{center}
$ ds^2 = -N^2(t) dt^2 + a^2(t) (dx^2 + dy^2 + dz^2) $. 
\end{center}
To investigate the evolution of these spacetimes within the framework of $ f(Q) $ gravity (see references [23–26] for more details), we must consider the corresponding non-metricity scalar for the above metric, which is expressed as:
\begin{center}
$ Q = \frac{6 H^2}{N^2} $.
\end{center}
Following the approach outlined in references [23, 24], we leverage the flexibility of $ f(Q) $ theories that permit specific choices of the lapse function $ N $. This is feasible because $ Q $ retains a residual time-reparametrization invariance, notwithstanding certain theoretical caveats. Consequently, we can fix $ N(t) = 1 $ for our analysis. In order to obtain the field equations, we simplify the equation by assuming that $ 8\pi G=1 $. We also impose the splitting condition $ f(Q)=Q+F(Q) $ and utilise the FLRW metric. As a result, we derive the related field equations as \cite{Khyllep:2021wjd, BeltranJimenez:2017tkd, BeltranJimenez:2019tme, Anagnostopoulos:2022gej,  Vishwakarma:2023brw},
\begin{equation}
3H^{2}= \rho + \frac{F}{2}-QF_{Q},
\end{equation}
\begin{equation}
(2QF_{QQ}+F_{Q}+1)\dot{H}+\frac{1}{4}(Q+2QF_{Q}-F)=-2p,
\end{equation}
where $ \rho $ and $ p $ denote the energy density and pressure, respectively. In this context, $ F_{Q} = \frac{dF}{dQ} $ and $ F_{QQ} = \frac{d^2 F}{dQ^2} $. These equations characterize the evolution of the Hubble parameter $H$ and the effective energy density $ \rho $, incorporating contributions from the non-metricity scalar $ Q $ and the function $ F(Q) $. This formulation offers a comprehensive framework for understanding the dynamics of the universe within the context of $ f(Q) $ gravity.

One can easily verify that the equations equations (6) and (7) are satisfying the standard conservation equation as stated below,
\begin{equation}
\dot{\rho}+3H \rho=0.
\end{equation}
From equation (6), we have
\begin{center}
$ 1=\frac{\rho}{3H^{2}} +\frac{\frac{F}{2}-QF_{Q}}{3H^{2}} $.
\end{center}
Hence the Friedmann's equation(6) can be simply written as
\begin{equation}
\Omega_{\mathrm{m}}+\Omega_{\mathrm{de}}=1 ,
\end{equation}
where \begin{equation}
\Omega_{\mathrm{m}}=\frac{\rho}{3H^{2}} , 
\end{equation}
and \begin{equation}
\Omega_{\mathrm{de}}=\frac{\frac{F}{2}-QF_{Q}}{3H^{2}}.
\end{equation}

It is possible to define the field equations of $F(Q)$ gravity in the dark energy sector ($ p_{\mathrm{de}} $ and $ \rho_{\mathrm{de}} $) as follows:
\begin{equation}
\rho_{\mathrm{de}}=\frac{1}{16\pi G}\left[ F-2QF_{Q} \right] ,
\end{equation}
\begin{equation}
p_{\mathrm{de}}=\frac{1}{16\pi G}\left[ 4(F_{Q}+2QF_{QQ})\dot{H} -F+ 2QF_{Q}\right].
\end{equation}
 Substituting the non-metricity Scalar $ Q=6H^{2} $, the EoS parameter of the dark energy sector ($ \displaystyle w_{\mathrm{de}}$) can be obtained as, 
 \begin{equation}
  \displaystyle w_{\mathrm{de}}= -1+ \frac{4(F_{Q}+2QF_{QQ}) }{F-2QF_{Q}}\frac{\dot{H}}{H^{2}}.  
 \end{equation}
Furthermore, the total EoS ($ \displaystyle w_{\mathrm{tot}}$) and the deceleration parameter ($ q $) can be defined as,
\begin{equation}
\displaystyle w_{\mathrm{tot}}= -1- \frac{2\dot{H}}{3 H^{2}}\equiv \frac{p_{\mathrm{m}}+p_{\mathrm{de}}}{\rho_{\mathrm{m}}+\rho_{\mathrm{de}}}, 
\end{equation}
\begin{equation}
q= -1 -\frac{\dot{H}}{H^{2}}.
\end{equation}

The matter density contrast, $\delta = \frac{\delta_{\rho}}{\rho} $, signifies the perturbation in the matter energy density, and is the primary focus of linear perturbation analysis. In the quasi-static regime, the following equation governs the evolution of matter density contrast, as described in \cite{Khyllep:2022spx}:
\begin{equation}
\ddot{\delta} + 2H\dot{\delta} = \frac{\rho \delta}{2(1 + F_{Q})},
\end{equation}
where the denominator on the right-hand side signifies the effective Newtonian constant. It is noteworthy that in the context of scales much smaller than the cosmic horizon, temporal derivative terms in the perturbation equations are often neglected. This simplification results in a formulation primarily governed by spatial derivative terms \cite{Anagnostopoulos:2022gej, Lazkoz:2019sjl}.

\section{Center Manifold Theory}
Central Manifold Theory (CMT) is a specialized area within dynamical systems theory, focused on examining the behavior of systems in the vicinity of fixed points. The foundational mathematical framework of CMT was comprehensively outlined by Perko \cite{Perko}. Traditional linear stability theory often falls short in accurately describing the stability of critical points when the associated eigenvalues include zero. In contrast, CMT enables a stability analysis by effectively reducing the system's dimensionality in the neighborhood of these points. As a system traverses a critical point, its behavior is governed by the invariant local center manifold, denoted as $W_{c}$. This central manifold $W_{c}$ is associated with eigenvalues possessing zero real parts, and the dynamics within this manifold encapsulate the key characteristics of the system's behavior near equilibrium \cite{Bahamonde:2017ize, Tamanini:2014nvd}.

Consider the dynamical system defined by
\begin{center}
\textbf{$ \varsigma' = F(\varsigma) $}
\end{center}
where $ \varsigma = (\mu, \nu) $. A geometrical space is deemed a center manifold for this system if it can be locally expressed as:
\begin{center}
$W_{c} = \left\{ (\mu, \nu) \in \mathbb{R} \times \mathbb{R} : \nu = h(\mu), \, |\mu| < \delta, \, h(0) = 0, \, \nabla h(0) = 0 \right\} $
\end{center}
for a sufficiently small $\delta$, where $ h(\mu) $ is a sufficiently regular function on $\mathbb{R}$.

\begin{flushleft}
\textbf{Definition 1} (Stable Fixed Point)
\end{flushleft}
A fixed point $ \varsigma_{0} $ of the system \textbf{$ \varsigma'=F(\varsigma) $} is stable if for every $\varepsilon>0$, we can find a $\delta$ such that for any solution $ \eta(t) $ of system \textbf{$ \varsigma'=F(\varsigma) $} satisfying $\Vert\eta(t_{0})-\varsigma_{0}\Vert<\delta$, then the solution $\eta(t)$ exists for all $t\geq t_{0}$ and it will satisfy $\Vert\eta(t)-\varsigma_{0}\Vert<\varepsilon$ for all $t\geq t_{0}$.

In simple words, a fixed point $\varsigma_{0}$ within a system described by $\varsigma' = F(\varsigma)$ is stable if all solutions $\varsigma(t)$, which begin close to $\varsigma_{0}$, stay in the vicinity of this point as time progresses. In essence, it means that nearby points gravitate towards and remain near $\varsigma_{0}$ over time.

\begin{flushleft}
\textbf{Definition 2} (Asymptotically Stable Fixed Point)
\end{flushleft}
A fixed point $ \varsigma_{0} $ of the system \textbf{$ \varsigma'=F(\varsigma) $} is called asymptotically stable if for every $\varepsilon>0$, we can find a $\delta$ such that for any solution $ \eta(t) $ of system \textbf{$ \varsigma'=F(\varsigma) $} satisfying $\Vert\eta(t_{0})-\varsigma_{0}\Vert<\delta$, then $\lim_{\mu\rightarrow\infty}\eta(t)= \varsigma_{0}$.

In another way, a fixed point $\varsigma_{0}$ in the system $\varsigma' = F(\varsigma)$ is considered asymptotically stable if a system is both stable and its perturbations from its equilibrium state gradually approach zero over time, it is referred to as``asymptotically stable." It indicates that after being perturbed, the system not only returns to its equilibrium state but also converges towards it, with the deviations decreasing as time approaches infinity. 

The Center Manifold Theory (CMT) analysis is conducted through the following steps:
\begin{enumerate}
\item \textbf{Coordinate Translation:} Initially, the coordinates of the non-hyperbolic critical points are translated to the origin, resulting in a set of autonomous equations expressed in the new coordinate system.

\item \textbf{Reformulation of the Dynamical System:} The transformed dynamical system is then expressed in the standard form, facilitating further analysis, given by.
\begin{equation}
\varsigma' = \begin{pmatrix}
\mu'  \\
\nu'
\end{pmatrix} = \begin{pmatrix}
A \mu  \\
B \nu
\end{pmatrix} + \begin{pmatrix}
\varphi(\mu, \nu)  \\
\psi(\mu, \nu)
\end{pmatrix},
\end{equation}
where the functions $\varphi$ and $\psi$ meet the following requirements:
\begin{center}
$\varphi(0,0) = 0$ and $\nabla \varphi(0,0) = 0$,
\end{center}
\begin{center}
$\psi(0,0) = 0$ and $\nabla \psi(0,0) = 0$.
\end{center}
The symbol $\nabla$ represents the gradient operator. Within this system, $A$ and $B$ are square matrices, each possessing eigenvalues with real portions equal to zero and negative, respectively. 

\item \textbf{Determining the Function $ h(\upsilon) $:}
After that, a function $h(\upsilon) $ is found, usually with the help of a series expansion that incorporates a $\upsilon^{2} $ term. This function $ h(\upsilon) $ satisfies the following quasilinear partial differential equation:
   \begin{equation*}
   \begin{split}
   \mathcal{N} h(\upsilon)  \equiv  \nabla h(\upsilon) \left[ A\upsilon + f(\upsilon, h(\upsilon)) \right. & \\ 
   \left. - Bh(\upsilon) - g(\upsilon, h(\upsilon)) \right] = 0,
   \end{split}
   \end{equation*}

   with the conditions $ h(0) = 0 $ and $ \nabla h(0) = 0 $.

\item \textbf{Dynamics on the Center Manifold:} By substituting the approximated solution of $ h(\upsilon) $ obtained from the previous equation, the dynamics of the original system restricted to the center manifold is given by:
\begin{equation}
\upsilon'=A\upsilon+\varphi(\upsilon,h(\upsilon)),
\end{equation}
for $ \upsilon\in \Bbb R $ is sufficiently small.

\item \textbf{Final Form of the Reduced System:} The equation $ \upsilon'=A\upsilon+\varphi(\upsilon,h(\upsilon) $ is further reduced to the form $ \upsilon' = k\upsilon^{n} $, where $ k $ represents a constant value and $ n $ denotes a positive integer, specifically referring to the term with the lowest order in the series expansion.
\begin{itemize}
\item If $ k < 0 $ and $ n $ is an odd integer,  it can be inferred that the system is stable, which consequently implies the stability of the original system.
\item  Under all other circumstances, both the reduced system and the original system will display instability.
\end{itemize}
\end{enumerate}

\section{Dynamical System Analysis}
In this section, we build a dynamical system based on the background and perturbed equations of a generic function $ F(Q)$. This is accomplished by transforming equations (6), (7) and (17) into first-order autonomous systems, denoted as:
\begin{equation}
 x = \frac{F}{6 H^{2}}, \quad y = -2 F_{Q}, \quad \xi = \frac{d(\ln \delta)}{d(\ln a)}.
\end{equation}
In this case, the variable $\xi $ tracks the expansion of matter disturbances, whereas the variables $ x $ and $ y $ are linked to the evolutionary dynamics of the background of the cosmos. The matter density contrast is thus positive in all time. The identification of perturbations in matter is as an increase when $\xi > 0 $ and a decrease when $\xi < 0 $.

The cosmic background parameters, specifically $\Omega_{\mathrm{m}}$, $\Omega_{\mathrm{de}}$, and $ \displaystyle w_{\mathrm{de}}$, are defined by the following expressions:
\begin{align}
\Omega_{\mathrm{m}} &= 1 - x - y, \\
\Omega_{\mathrm{de}} &= x + y, \\
\displaystyle w_{\mathrm{de}} &= -1 + \frac{(4QF_{QQ}-y)}{3(x+y)} \frac{\dot{H}}{H^{2}}.
\end{align}
These parameters enable the reformulation of the cosmological equations into a dynamical system, utilizing the variables defined in equation (20), as follows:
\begin{align}
x' &= -\frac{\dot{H}}{H^{2}}(y+2x), \\
y' &= -\frac{\dot{H}}{H^{2}}4QF_{QQ}, \\
\xi' &= -\xi(\xi+2)+\frac{3(1-x-y)}{2-y}-\frac{\dot{H}}{H^{2}}\xi,
\end{align}
where $ (') $ denotes differentiation with respect to $ \ln a $ and $ (.) $ denotes differentiation with respect to $ t $. Additionally, the following relation holds:
\begin{equation}
\frac{\dot{H}}{H^{2}} = \frac{3(x+y-1)}{4QF_{QQ} - y + 2}.
\end{equation}

The perturbed space, $\mathbb{P} $, which contains the variable $\xi $, and the background phase space, $\mathbb{B} $, which includes the variables $x $ and $y $, are both components of the composite space that constitutes the physical system. This system's combined phase space is defined as follows, given the physical condition $0 \leq \Omega_{\mathrm{m}} \leq 1 $:
\[
\Psi = \mathbb{B} \times \mathbb{P} = (x, y, \xi) \in \mathbb{R}^{2} \times \mathbb{R} : 0 \leq x + y \leq 1.
\]
Crucially, when orbits from product space $\Psi $ are projected onto background space $\mathbb{B} $, the resulting reduction to the matching orbits in the background space needs to be observed.

The key to unravelling the system's dynamical evolution is in pinpointing and assessing its critical points. When $\xi > 0 $, the system becomes unstable in the presence of matter perturbations, suggesting that these disturbances can grow infinitely. In contrast, the decay of matter disturbances is reflected by a stable point with $ \xi \ 0 $, which indicates the system's asymptotic stability with regard to perturbations. It follows that matter perturbations stay constant when the system attains a stable point with $\xi = 0 $. It should be emphasised that the growth of matter perturbations, especially at unstable or saddle locations where $\xi > 0 $, does not always indicate a stable condition. The analysis of the universe's matter-dominated period requires this comprehension. A stable late-time attractor, denoted by $ \xi = 0 $, which indicates an acceleration phase, must precede such unstable or saddle points \cite{BeltranJimenez:2019tme}.

Determining the function $ F(Q) $ and then establishing the term $ QF_{QQ} $ are crucial for an exhaustive examination. Three particular models, highlighted for their noteworthy cosmic phenomenology in Sec. IV, will be easier to investigate in this way. The next sections will provide a more in-depth explanation of these models.

\subsection{MODEL I : $ f(Q)=\frac{\alpha '}{2} \sqrt{Q}\ln(\gamma ' Q)+\beta Q $}

The class of models we will investigate is based on the form of f(Q) given by Refs. \cite{Anagnostopoulos:2022gej, Ayuso:2021vtj, Goncalves:2024sem}
\begin{equation}
f(Q) = \frac{\alpha '}{2}\sqrt{Q} \ln(\gamma ' Q)+ \beta Q,
\end{equation}
where the parameters $\alpha'$, $\beta$, and $\gamma'$, are constants. It is important to note that the equivalent of General Relativity (GR) is recovered by choosing $\alpha' = 0$ and $\beta = 1$. While the constant $\beta$ is dimensionless, the dimensions for the other two parameters are such that $[\alpha'] = [Q]^{1/2}$ and $[\gamma'] = [Q]^{-1}$. Defining the following dimensionless constant parameters:
\begin{eqnarray*}
\alpha \equiv \frac{\alpha'}{\sqrt{Q_{0}}}, \\
\gamma \equiv \gamma'Q_{0},
\end{eqnarray*}
where the constant $Q_{0}$ is the present value of $Q$, i.e., $Q_0 \equiv 6H_{0}^{2}$. Consequently, equation (19) becomes:
\begin{equation*}
f(Q) = \frac{\alpha}{2}\sqrt{Q_{0} Q} \ln \left( \gamma \frac{Q}{Q_{0}} \right) + \beta Q,
\end{equation*}
which is the form we will use throughout this work.

The given $ f(Q) $ model is motivated by the need to explain cosmic acceleration without invoking a cosmological constant or exotic dark energy. It extends General Relativity (GR) by introducing a logarithmic correction term, $ \sqrt{Q} \ln(Q) $, which is inspired by quantum corrections, renormalization group flows, and entropic gravity models. The model ensures that GR is recovered for $ \alpha' = 0 $ and $ \beta = 1 $, while the logarithmic modification introduces scale-dependent deviations that become significant at late times, potentially driving cosmic acceleration. By defining dimensionless parameters using the present value of $ Q $, the formulation avoids fine-tuning issues and maintains scale invariance. The model also has implications for structure formation, modifying the growth of perturbations and potentially offering observable deviations from standard $\Lambda$CDM predictions.

Imposing the splitting $ f(Q)=Q+F(Q) $ the above equation becomes,
\begin{equation}
F(Q)=\frac{\alpha}{2} \sqrt{QQ_{0}}\ln(\gamma \frac{Q}{Q_{0}})+\beta Q - Q, 
\end{equation}
Specifically, when evaluating the second derivative of $F(Q)$ with respect to $ Q $ ($QF_{QQ}$), it follows that 
\begin{center}
$ QF_{QQ}=\frac{-x}{4} $,
\end{center}
Therefore the autonomous systems given by equations (24)-(27) becomes,
\begin{align}
x' &= \frac{3(x+y-1)}{x+y-2}(y+2x), \\
y' &=  \frac{3(x+y-1)}{x+y-2}(-x), \\
\xi' &= -\xi(\xi+2) + \frac{3(1-x-y)}{2-y} +  \frac{3(x+y-1)}{x+y-2}\xi.
\end{align}
The singularities of the given system occur at $ y = 2 $ and $ x = 0 $,
and the EoS parameters and deceleration parameter becomes,
\begin{align}
\displaystyle w_{\mathrm{de}} &= -1 + \frac{(x-2y)(x+y-1)}{2(x+y)(2-x-y)}, \\
\displaystyle w_{\mathrm{tot}} &= -1 + \frac{2(x+y-1)}{(x+y-2)}, \\
q &= -1 + \frac{3(x+y-1)}{(x+y-2)}. 
\end{align}

Four critical points have been identified and are detailed in Table I, along with their associated cosmological characteristics. Table II presents the eigenvalues of the Jacobian matrix for both background and perturbation levels.

\begin{table*}
\caption{Critical points (CP), matter, dark energy, perturbation and EoS parameters}
\begin{tabular}{ ||p{1.5cm}||p{2.6cm}||p{0.8cm}||p{0.8cm}||p{0.8cm}||p{0.8cm}||p{0.8cm}||p{0.8cm}|| }
 \hline
 \textbf{CP} & $\mathbf{(x_{c}, y_{c}, \xi_{c})}$ & $\mathbf{\Omega_{\mathrm{m}}}$ & $\mathbf{\Omega_{\mathrm{de}}}$ & $ \mathbf {\xi} $ & $ \mathbf{\displaystyle w_{\mathrm{de}}} $ & $\mathbf{ \displaystyle w_{\mathrm{tot}}} $ & $ \mathbf{q}$ \\
 \hline
 $ A_{1} $ & $ (1-y,y,0) $ & $ 0 $ & $ 1 $ & $ 0 $ & $ -1 $ & $ -1 $ & $ -1 $  \\
 \hline
 $ A_{2} $ & $ (1-y,y,-2) $ & $ 0 $ & $ 1 $ & $ -2 $ & $ -1 $ & $ -1 $ & $ -1 $  \\
 \hline
  $ A_{3} $ & $ (0,0,-\frac{3}{2}) $ & $ 1 $ & $ 0 $ & $ -\frac{3}{2} $ & $ --- $ & $ 0 $ & $ \frac{1}{2} $ \\
 \hline
  $ A_{4} $ & $ (0,0,1) $ & $ 1 $ & $ 0 $ & $ 1 $ & $ --- $ & $ 0 $ & $ \frac{1}{2} $ \\
 \hline
\end{tabular}

\label{1}
\end{table*} 

\begin{table}
\caption{Eigen values and Stability conditions}
\begin{tabular}{ ||p{1.5cm}||p{2.4cm}||p{3.2cm}|| }
 \hline
 \textbf{CP} & \textbf{Eigen-Values} & \textbf{Stability Condition}  \\
 \hline
 $ A_{1} $ & $ (0,-3,-2) $ & Stable  \\
 \hline
 $ A_{2} $ & $ (0,-3,2) $ & Saddle  \\
 \hline
  $ A_{3} $ & $ (\frac{5}{2},\frac{3}{2},\frac{3}{2}) $ & Unstable \\
 \hline
  $ A_{4} $ & $ (-\frac{5}{2},\frac{3}{2},\frac{3}{2}) $ & Saddle \\
 \hline
\end{tabular}

\label{2}
\end{table} 

 \vspace{5mm} \textbf{* Critical Point $A_{1}=(1-y,y,0)$ :}
At this critical point, the system exhibits the following eigenvalues: $(0, -3, -2)$. This scenario represents a de Sitter universe dominated entirely by dark energy with no contribution from matter i.e $\Omega_{\mathrm{m}} = 0$ and $\Omega_{\mathrm{de}} = 1$. The values of the equation of state parameters, $\displaystyle w_{\mathrm{de}}=-1 $, $\displaystyle w_{\mathrm{tot}}=-1 $ and the deceleration parameter $q=-1 $ indicate an accelerated expansion consistent with a cosmological constant-like behavior. The eigenvalues suggest that the critical point $A_{1}$ is a stable attractor in the phase space of the dynamical system, with one zero eigenvalue corresponding to a marginal stability direction and the other two negative eigenvalues indicating stability in their respective directions. At the perturbation level, $ \xi = 0 $ indicates that matter perturbations are constant.The one-dimensional equivalent curve has one eigenvalue that vanishes and two other eigenvalues are $-3$ and $-2$ that, according to the non-vanishing eigenvalues, are stable. The fact that the Jacobian matrix has a zero eigenvalue further confirms that the critical point $A_{1} $ is non hyperbolic. The system of equations shows asymptotic stability at this equilibrium point when the centre manifold theory is used.

The central variable in this context is $x$, and the stable variables are $(y, \xi)$. The corresponding matrices $A$ and $B$ are characterized by $A = 0$ and $B = \begin{pmatrix}
-3 & 0\\
0 & -2
\end{pmatrix}$. The structure of the center manifold takes the form $y = h_{1}(x)$ and $\xi = h_{2}(x)$, with the approximation $N$ comprising two components.
\begin{flushleft}
$ N_{1}(h_{1}(x))=h_{1}'(x)\frac{3(x+h_{1}(x)-1)}{x+h_{1}(x)-2}(h_{1}(x)+2x)+\frac{3x(x+h_{1}(x)-1)}{x+h_{1}(x)-2} $,
\end{flushleft}
\begin{flushleft}
$ N_{2}(h_{2}(x))=h_{2}'(x) \frac{3(x+h_{1}(x)-1)}{x+h_{1}(x)-2}(h_{1}(x)+2x)+h_{2}(x)(h_{2}(x)+2) - \frac{3(1-x-h_{1}(x))}{2-h_{1}(x)} -  \frac{3(x+h_{1}(x)-1)}{x+h_{1}(x)-2}h_{2}(x)$.
\end{flushleft}
 \vspace{5mm} \textbf{For zeroth approximation:}\\ $ N_{1}(h_{1}(x))=\frac{-3x}{x-2}+\mathcal{O}(x^{2}) $ and\\ $ N_{2}(h_{2}(x))=\frac{3(1-x)}{2}+\mathcal{O}(x^{2}) $.
\\Therefore the reduced equation gives us
\begin{flushleft}
$ x'=\frac{-21x}{6x-4}+O(x^{2}) $
\end{flushleft}
The results show that for every value of $x$, the linear behaviour is always negative. At the equilibrium point, the system of equations (30)–(32) shows asymptotic stability according to the central manifold theory.

Finally, at both the background and perturbation levels, the discussion highlights that the late universe is predominantly influenced by dark energy.
 
\vspace{5mm} \textbf{* Critical Point $ A_{2}=(1-y,y,-2) $: }  
The critical point $A_{2} = (1-y, y, -2)$ signifies a state where the universe is completely dominated by dark energy ($\Omega_{\mathrm{de}} = 1$) with no contribution from matter ($\Omega_{\mathrm{m}} = 0$). The values $\displaystyle w_{\mathrm{de}} = -1$ and $\displaystyle w_{\mathrm{tot}} = -1$ correspond to a cosmological constant-like dark energy, leading to a de Sitter expansion characterized by a deceleration parameter $q = -1$. This critical point can thus be interpreted as a de Sitter attractor solution, indicating that the universe undergoes accelerated expansion driven solely by dark energy. The eigenvalues associated with the critical point $A_{2}$ are $(0, -3, 2)$, indicating a saddle point configuration. At the perturbation level, the critical point $A_{2}$ does not correlate to a universe dominated by late-time dark energy, in contrast to $A_{1}$. Its saddle shape and negative values for $\displaystyle w_{\mathrm{de}} $ and $\displaystyle w_{\mathrm{tot}} $ represent that the universe is going through an inflationary period.

\vspace{5mm} \textbf{* Critical Point $ A_{3}=(0,0,-\frac{3}{2}) $: }
The critical point $A_{3} = (0, 0, -\frac{3}{2})$ represents a universe dominated by matter ($\Omega_{\mathrm{m}} = 1$) with no contribution from dark energy ($\Omega_{\mathrm{de}} = 0$). The perturbation parameter $\xi = -\frac{3}{2}$ is fixed at this point. Since dark energy is absent, the equation of state parameter for dark energy ($\displaystyle w_{\mathrm{de}}$) is not defined. The total equation of state parameter $\displaystyle w_{\mathrm{tot}}= 0$ indicates a matter-dominated universe, which corresponds to a deceleration parameter $q = \frac{1}{2}$. The eigenvalues associated with this critical point are $\left(\frac{5}{2}, \frac{3}{2}, \frac{3}{2}\right)$. Since all the eigenvalues are positive, the critical point $ A_{3} $ is generally unstable. Also, $ \xi=-\frac{3}{2} $, suggests the decay in matter perturbation. The critical point $A_{3} = (0, 0, -\frac{3}{2})$ represents a matter-dominated universe with no dark energy contribution. The stability analysis via center manifold theory reveals that this critical point has unstable directions due to the presence of positive eigenvalues $\left(\frac{5}{2}, \frac{3}{2}, \frac{3}{2}\right)$. Consequently, the critical point $A_{3}$ is generally unstable.

\vspace{5mm} \textbf{* Critical Point $ A_{4}=(0,0,1) $: } 
The critical point $A_{4} = (0, 0, 1)$ signifies a universe predominantly governed by matter ($\Omega_{\mathrm{m}} = 1$) with no contribution from dark energy ($\Omega_{\mathrm{de}} = 0$). The perturbation parameter $\xi = 1$, indicates growth in matter perturbation. As there is no dark energy contribution, the equation of state parameter for dark energy ($\displaystyle w_{\mathrm{de}}$) remains undefined. The total equation of state parameter $\displaystyle w_{\mathrm{tot}} = 0$ indicates a matter-dominated universe, corresponding to a deceleration parameter $q = \frac{1}{2}$.

The eigenvalues $\left(-\frac{5}{2}, \frac{3}{2}, \frac{3}{2}\right)$ reveals that the critical point exhibits both stable and unstable directions. In particular, eigenvalues that are negative suggest stability in one direction whereas eigenvalues that are positive indicate instability in other directions. Therefore, $A_{4}$ is typically susceptible to saddle instability. After diverging from this point, trajectories converge on a late-time stable point. This finding implies that this specific critical point is a key decision for understanding structure development during the matter-dominated period, which successfully handles dynamics at both the background and perturbation levels. The critical point $A_{4} = (0, 0, 1)$ represents a matter-dominated universe with no dark energy contribution. Hence, $ A_{4} $ may be the point, which can explain how structures are formed, both at background and perturbation level.

\begin{figure}
    \centering
    \includegraphics[scale=0.53]{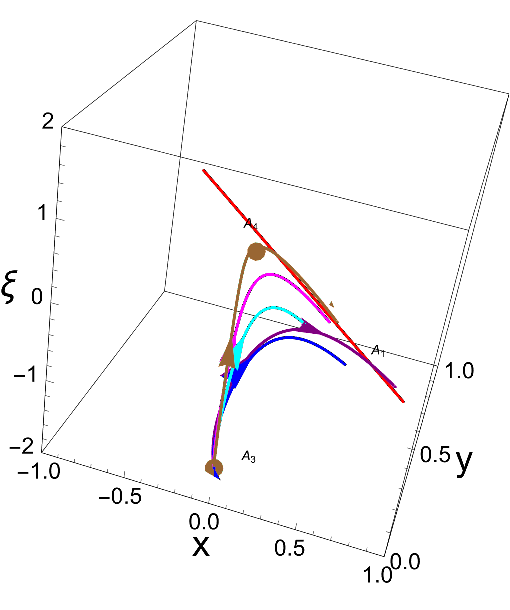}
    \caption{3D phase  portrait for \textbf{Model-I}.}
\end{figure}

The analysis reveals the existence of two dark energy-dominated critical points (($A_{1}$ and $A_{2}$) and two matter-dominated critical points ($A_{3}$ and $A_{4}$) within the framework of the logarithmic form of $f(Q)$ gravity. The matter-dominated points, ($A_{3}$ and $A_{4}$, are found to be inherently unstable. Specifically, critical point $A_{3}$, which functions as a saddle point, is indicative of a defined growth rate in matter perturbations. In contrast, critical point $A_{4}$, identified as an unstable node, signifies the decay of matter perturbations. Notably, $A_{4}$ exhibits accelerated expansion behavior at the background level only. On the other hand, critical point $A_{1}$ demonstrates this accelerated expansion behavior consistently at both the background and perturbation levels, distinguishing it as a stable configuration.

Figure 1 presents the phase portrait within a three-dimensional space, illustrating the evolution of the selected trajectory as it transitions from matter-dominated to dark-energy-dominated critical points. The diagram clearly depicts the sequential progression of the trajectory, transitioning from the unstable node at critical point $A_{3}$ to the saddle instability at $A_{4}$, and finally settling at the stable node represented by critical point $A_{1}$.

\begin{figure} 
   \centering 
   \mbox{\includegraphics[scale=0.53]{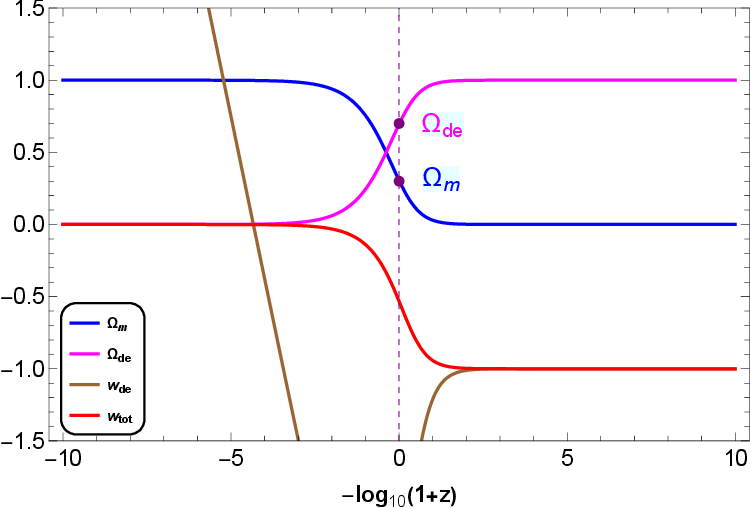}}   
    \hspace{10px}
    \mbox{\includegraphics[scale=0.53]{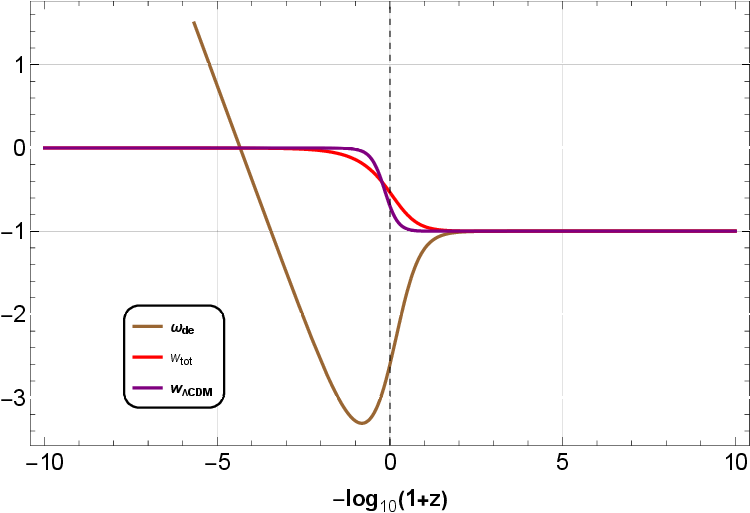}}
    \hspace{10px}
    \mbox{\includegraphics[scale=0.53]{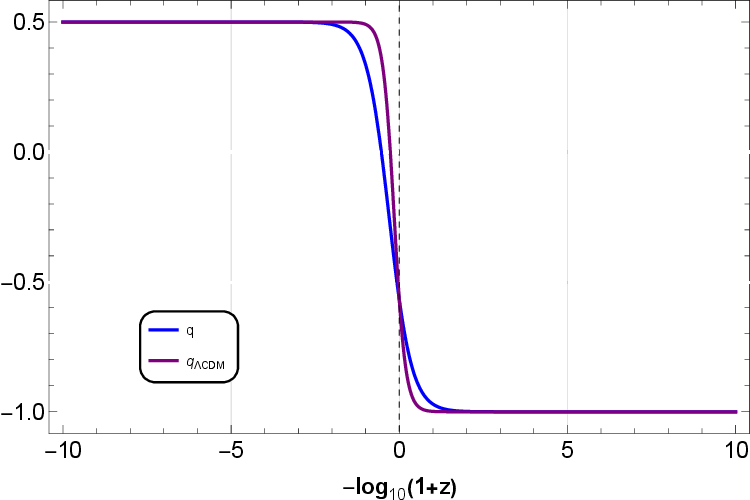}}
    \caption{ The evolution of the density parameters (shown in the \textbf{Upper panel}), the EoS parameters (illustrated in the \textbf{Middle panel}), and the deceleration parameter (depicted in the \textbf{Lower panel}) are presented for \textbf{Model-III}. The initial conditions are set as $x = 10^{-2}$, $y = 10^{-6}$. The vertical dashed line represents the present time.}
   \label{Phase Portrait for the dynamical system of Model I}
\end{figure}

Figure 2 (Upper and Middle panels) illustrates the evolutionary history of the density parameter and equation of state (EoS) parameter. In Uppere panel, the universe undergoes a transition from a matter-dominated phase to an accelerated expansion era at the late times. The current density parameters for the matter and dark energy sectors are approximately $ \Omega_{\mathrm{m}} \approx 0.3 $ and $\Omega_{\mathrm{de}} \approx 0.7$, respectively. The Middle panel shows the the total EoS parameter, which begins in a matter dominated era ($ \displaystyle w_{\mathrm{tot}}=0 $) and gradually transitions to the dark energy sector ($ \displaystyle w_{\mathrm{tot}} \approx -1 $) as the universe evolves. Concurrently, the dark energy EoS parameter approached $ -1 $ in the late stages of evolution, with the present value of $ \displaystyle w_{\mathrm{de}}=-1 $ aligniing with the current observational range of $ \displaystyle w_{\mathrm{de}}= -1.028 \pm 0.032 $ \cite{Planck:2018fzr}. In the Lower panel, the deceleration parameter exhiboits a transition from a decelerating to an accelerating phase, with the transition point accurring at $ z= 0.59 $ and the current deceleration parameter value being $ q_{0}= -0.57 $ \cite{Camarena:2019moy}.

\subsection{MODEL II : $ f(Q)=Q-6 \lambda M^{2} \left(  \frac{Q}{6 M^{2}} \right) ^{\alpha}  $}
The polynomial model, as introduced in Refs. \cite{BeltranJimenez:2019tme, Anagnostopoulos:2022gej} given by
\begin{equation}
f(Q)=Q-6 \lambda M^{2} \left(  \frac{Q}{6 M^{2}} \right) ^{\alpha}
\end{equation} serves as a generalization of the square-root model. While the square-root model introduces modifications to the evolution of perturbations without altering the standard General Relativity (GR) background dynamics, the polynomial model extends these modifications to both the background and perturbation levels. The formulation of the polynomial model is expressed as follows:
\begin{equation}
F(Q)=-6 \lambda M^{2} \left(  \frac{Q}{6 M^{2}} \right) ^{\alpha} ,
\end{equation}
In this context, $\lambda$ and $\alpha$ are dimensionless parameters, while $M$ represents a mass scale, typically of the order of $\sqrt{\Lambda}$, where $\Lambda$ denotes the standard cosmological constant. When $\alpha = 0$, the theory reduces to the symmetric teleparallel equivalent of General Relativity (STEGR) with an additional cosmological constant term equal to $6\lambda M^2$. For $\alpha = 1$, the theory corresponds to STEGR with a modified gravitational constant, $G \rightarrow G/(1-\lambda)$. More generally, values of $\alpha > 1$ are predominantly relevant to early Universe cosmology, while $\alpha < 1$ is pertinent to the description of dark energy, making it significant for late-time cosmological evolution. Additionally, in the latter case, the theory exhibits an asymptotic approach to General Relativity in the early Universe. It is also noteworthy that the case $\alpha = -1$ has been tested against late Universe observations, as discussed in Ref. \cite{Ayuso:2020dcu}.

The polynomial $ f(Q) $ model generalizes GR by modifying both background and perturbation evolution, making it relevant for both early- and late-time cosmology. Unlike the square-root model, which primarily affects perturbations while preserving the GR background, this model introduces modifications at both the background and perturbation levels, providing a more comprehensive framework for studying deviations from GR. It reduces to STEGR with an effective cosmological constant for $ \alpha = 0 $ and modifies the gravitational constant for $ \alpha = 1 $. Larger values of $ \alpha $ impact early Universe dynamics, while smaller values ($\alpha < 1$) describe late-time acceleration, offering a dark energy alternative. The model smoothly transitions to GR at high energies and has been tested for observational viability, making it a flexible extension of standard cosmology.

As
\begin{center}
$ QF_{QQ}=(1-\alpha)\frac{y}{2} $,
\end{center}
The system (24)--(27) becomes
\begin{align}
x'&=\frac{3(1-x-y)(y+2x)}{y(1-2\alpha)+2}, \\
y'&=\frac{6y(1-x-y)(1-\alpha)}{y(1-2\alpha)+2}, \\
\xi' &= -\xi(\xi+2) + \frac{3(1-x-y)}{2-y} + \frac{3(1-x-y)\xi}{y(1-2\alpha)+2}.
\end{align}
The singularities of the given system are located at $ y = 2 $ or $ y = -\frac{2}{1 - 2\alpha} $.
The corresponding Eos and deceleration parameters reduces to, 
\begin{align}
\displaystyle w_{\mathrm{de}} &= -1 - \frac{y(2-\alpha)(x+y-1)}{(x+y)(y(1-2\alpha)+2)}, \\
\displaystyle w_{\mathrm{tot}} &= -1 - \frac{2(x+y-1)}{y(1-2\alpha)+2}, \\
q &= -1 - \frac{3(x+y-1)}{y(1-2\alpha)+2}.
\end{align}

Four critical points have been identified and are detailed in Table III, which outlines their corresponding cosmological characteristics. Table IV provides the eigenvalues of the Jacobian matrix at both the background and perturbation levels.

\begin{table*}
\caption{Critical points (CP), matter, dark energy, perturbation and EoS parameters}
\begin{tabular}{ ||p{1.5cm}||p{2.6cm}||p{0.8cm}||p{0.8cm}||p{0.8cm}||p{0.8cm}||p{0.8cm}||p{0.8cm}|| }
 \hline
 \textbf{CP} & $\mathbf{(x_{c}, y_{c}, \xi_{c})}$ & $\mathbf{\Omega_{\mathrm{m}}}$ & $\mathbf{\Omega_{\mathrm{de}}}$ & $ \mathbf {\xi} $ & $ \mathbf{\displaystyle w_{\mathrm{de}}} $ & $\mathbf{ \displaystyle w_{\mathrm{tot}}} $ & $ \mathbf{q }$ \\
 \hline
 $ B_{1} $ & $ (1-y,y,0) $ & $ 0 $ & $ 1 $ & $ 0 $ & $ -1 $ & $ -1 $ & $ -1 $  \\
 \hline
 $ B_{2} $ & $ (1-y,y,-2) $ & $ 0 $ & $ 1 $ & $ -2 $ & $ -1 $ & $ -1 $ & $ -1 $  \\
 \hline
  $ B_{3} $ & $ (0,0,-\frac{3}{2}) $ & $ 1 $ & $ 0 $ & $ -\frac{3}{2} $ & $ --- $ & $ 0 $ & $ \frac{1}{2} $ \\
 \hline
  $ B_{4} $ & $ (0,0,1) $ & $ 1 $ & $ 0 $ & $ 1 $ & $ --- $ & $ 0 $ & $ \frac{1}{2} $ \\
 \hline
\end{tabular}

\label{1}
\end{table*} 

\begin{table}
\caption{Eigen values and Stability conditions}
\begin{tabular}{ ||p{1.5cm}||p{2.4cm}||p{3.2cm}|| }
 \hline
 \textbf{CP} & \textbf{Eigen-Values} & \textbf{Stability Condition}  \\
 \hline
 $ B_{1} $ & $ (0,-3,-2) $ & Stable  \\
 \hline
 $ B_{2} $ & $ (0,-3,2) $ & Saddle  \\
 \hline
  $ B_{3} $ & $ (3,\frac{29997}{10000},\frac{5}{2}) $ & Unstable \\
 \hline
  $ B_{4} $ & $ (3,\frac{29997}{10000},-\frac{5}{2}) $ & Saddle \\
 \hline
\end{tabular}

\label{2}
\end{table} 

 \vspace{5mm} \textbf{* Critical Point $B_{1}=(1-y,y,0)$ :}
The critical point $B_{1}$ corresponds to a state where the matter density parameter $\Omega_{\mathrm{m}} = 0$ and the dark energy density parameter $\Omega_{\mathrm{de}} = 1$. This indicates that the Universe is entirely dominated by dark energy, with no contribution from matter. Given that the equation of state parameters $ \displaystyle w_{\mathrm{de}} = -1 $ and $ \displaystyle w_{\mathrm{tot}} = -1 $, the critical point represents a de Sitter-like phase, characterized by a constant Hubble parameter and an accelerated expansion of the Universe. The deceleration parameter $ q = -1 $ further confirms this accelerated expansion, as it indicates a Universe that is expanding at an exponential rate.

The eigenvalues associated with this critical point are $(0, -3, -2)$. The presence of a zero eigenvalue suggests that the system has a marginally stable direction, meaning that small perturbations in this direction neither grow nor decay, leading to neutral stability. The negative eigenvalues, $-3$ and $-2$, indicate stability in the other directions, as perturbations in these directions will decay over time. Since the eigenvalues are of non-hyperbolic nature, we use center manifold theory for further analysis of the stability of the critical point  $ B_{1} $.

In this case, the stable variables are represented by $(y,\xi)$, and the central variable is $x$. The corresponding matrices $A$ and $B$ are characterized by $A = 0$ and $B = \begin{pmatrix}
-3 & 0\\
0 & -2
\end{pmatrix}$. The structure of the center manifold takes the form $y = h_{3}(x)$ and $\xi = h_{4}(x)$, with the approximation $N$ comprising two components.
\begin{flushleft}
$ N_{3}(h_{3}(x))=h_{3}'(x)\frac{3(1-x-h_{3}(x))(h_{3}(x)+2x)}{h_{3}(x)(1-2\alpha)+2}- \frac{6h_{3}(x)(1-x-h_{3}(x))(1-\alpha)}{h_{3}(x)(1-2\alpha)+2}$,
\end{flushleft}
\begin{flushleft}
$ N_{4}(h_{4}(x))=h_{4}'(x) \frac{3(1-x-h_{3}(x))(h_{3}(x)+2x)}{h_{3}(x)(1-2\alpha)+2} + h_{4}(x) (h_{4}(x)+2) - \frac{3(1-x-h_{3}(x))}{2-h_{3}(x)} - \frac{3(1-x-h_{3}(x))h_{4}(x)}{h_{3}(x)(1-2\alpha)+2} $.
\end{flushleft}
\vspace{5mm} \textbf{For zeroth approximation:}\\ $ N_{3}(h_{3}(x))=0+\mathcal{O}(x^{2}) $ and\\ $ N_{4}(h_{4}(x))=-\frac{3(1-x)}{2}+\mathcal{O}(x^{2}) $.
\\Therefore the reduced equation gives us
\begin{flushleft}
$ x'=-\frac{3}{2}+O(x^{2}) $
\end{flushleft}
which gives the negative linear part for all $ x $. Consequently, the system of equations (38)--(40) demonstrates asymptotic stability at the equilibrium point, consistent with the central manifold theory.
 
Overall, the critical point $B_{1}$ suggests a stable configuration where the Universe is dominated by dark energy, leading to a phase of accelerated expansion consistent with a de Sitter universe. The stability analysis through the eigenvalues implies that this state is robust against small perturbations, ensuring that the Universe will remain in this accelerated expansion phase if it reaches this critical point.

\vspace{5mm} \textbf{* Critical Point $ B_{2}=(1-y,y,-2) $: }  
The critical point $ B_{2}$ corresponds to a scenario where the matter density parameter $\Omega_{\mathrm{m}}$ is zero, indicating the absence of matter, and the dark energy density parameter $\Omega_{\mathrm{de}}$ is one, implying that the universe is entirely dominated by dark energy. The equation of state parameters are $ \displaystyle w_{\mathrm{de}} = -1 $ and $ \displaystyle w_{\mathrm{tot}} = -1 $, which are characteristic of a cosmological constant or vacuum energy. The deceleration parameter $ q = -1 $ suggests that the universe is undergoing accelerated expansion. The eigenvalues associated with this critical point are $(0, -3, 2)$, which indicate the stability of the point. Specifically, the presence of a zero eigenvalue suggests the existence of a center manifold, while the negative and positive eigenvalues imply a saddle-like behavior in the system's dynamics. The perturbation $\xi = -2$  represents a specific perturbative mode around this critical point. This combination of parameters indicates a universe in a de Sitter-like phase, where dark energy dominates, and the universe's expansion is accelerating. Nonetheless, owing to its saddle nature and the negative $\displaystyle w_{\mathrm{de}}$ and $ \displaystyle w_{\mathrm{tot}} $ value, this point portrays the inflationary epoch of the universe.

\vspace{5mm} \textbf{* Critical Point $ B_{3}=(0,0,-\frac{3}{2}) $: }
The critical point $B_{3}$ corresponds to a cosmological scenario where the universe is entirely dominated by matter, as indicated by $ \Omega_{\mathrm{m}} = 1 $ and $ \Omega_{\mathrm{de}} = 0 $. At this point, the equation of state parameter for dark energy, $ \displaystyle w_{\mathrm{de}} $, is not defined since dark energy is absent. The total equation of state parameter, $\displaystyle w_{\mathrm{tot}} = 0$, suggests that the universe behaves like a pressureless dust, consistent with a matter-dominated era. The deceleration parameter $ q = \frac{1}{2} $ implies that the universe is decelerating, as expected in a matter-dominated phase of cosmic evolution. This is consistent with the standard cosmological model during the period before dark energy becomes dominant. The perturbation parameter $ \xi = -\frac{3}{2} $ at this critical point indicates a specific perturbative behavior in the system.

The eigenvalues associated with this critical point are $ (3, \frac{29997}{10000}, \frac{5}{2}) $. The presence of all positive eigenvalues suggests that the critical point is an unstable node, indicating that small perturbations around this point will grow, leading the system away from this equilibrium. This instability reflects the transient nature of a matter-dominated universe, eventually giving way to other phases of cosmic evolution, such as dark energy domination.

\vspace{5mm} \textbf{* Critical Point $ B_{4}=(0,0,1) $: } 
The critical point $ B_{4} $ represents a scenario where the universe is entirely dominated by matter, with $\Omega_{\mathrm{m}} = 1$ and $\Omega_{\mathrm{de}} = 0$. In this configuration, $ \displaystyle w_{\mathrm{de}} $ is not defined because dark energy is absent in this scenario ($\Omega_{\mathrm{de}} = 0$). The total equation of state $\displaystyle w_{\mathrm{tot}} = 0 $, indicating that the universe behaves as a pressureless matter-dominated universe, which corresponds to non-relativistic matter. The deceleration parameter $ q = \frac{1}{2} $, which is characteristic of a universe dominated by matter. This value implies that the universe is decelerating in its expansion. The eigenvalues associated with this critical point are $(3, \frac{29997}{10000}, -\frac{5}{2})$. The positive eigenvalues suggest that the critical point is a saddle point in the phase space, indicating that it is unstable in certain directions whereas the negative eigenvalue indicates stability in at least one direction. The perturbation value is $\xi = 1 $, which corresponds to the specific perturbative analysis around this critical point.

Hence, the critical point $ B_{4} $ highlights a regime where the dynamics are dominated by matter, with dark energy playing no role. The instability suggested by the positive eigenvalues indicates that this state is not attractor-like, meaning the universe may not remain in this state indefinitely, potentially transitioning to another phase dominated by different components, such as dark energy, as the universe evolves.

\begin{figure}
    \centering
    \includegraphics[scale=0.53]{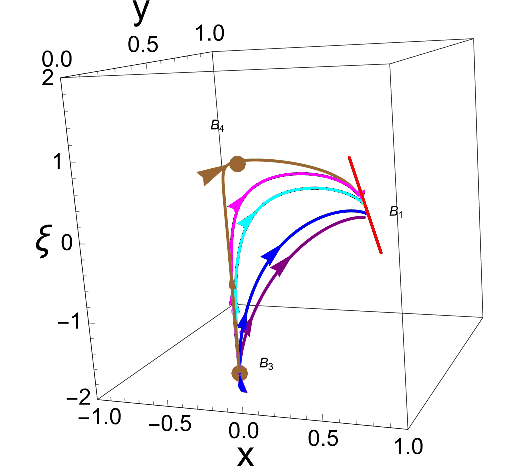}
    \caption{3D phase  portrait for \textbf{Model-II}.}
\end{figure}

The analysis identifies two dark energy-dominated critical points ($B_{1}$ and $B_{2}$) and two matter-dominated critical points ($B_{3}$ and $B_{4}$) within the context of the polynomial form of $f(Q) $ gravity. Both matter-dominated points, $B_{3}$ and $B_{4}$, are characterized by inherent instability. Specifically, critical point $B_{3}$, acting as a saddle point, reflects a defined growth rate in matter perturbations. Conversely, critical point $B_{4}$, identified as an unstable node, indicates the decay of matter perturbations. It is important to note that $B_{4}$ exhibits accelerated expansion behavior solely at the background level. In contrast, critical point $B_{1}$ consistently demonstrates this accelerated expansion behavior at both the background and perturbation levels, thereby establishing it as a stable configuration.

Figure 3 illustrates the phase portrait in three-dimensional space, depicting the trajectory's evolution as it transitions from matter-dominated to dark-energy-dominated critical points. The diagram clearly shows the sequential progression of the trajectory, beginning at the unstable node corresponding to critical point $B_{3}$, passing through the saddle instability at $B_{4}$, and ultimately stabilizing at the node represented by critical point $B_{1}$.

\begin{figure} 
   \centering 
   \mbox{\includegraphics[scale=0.53]{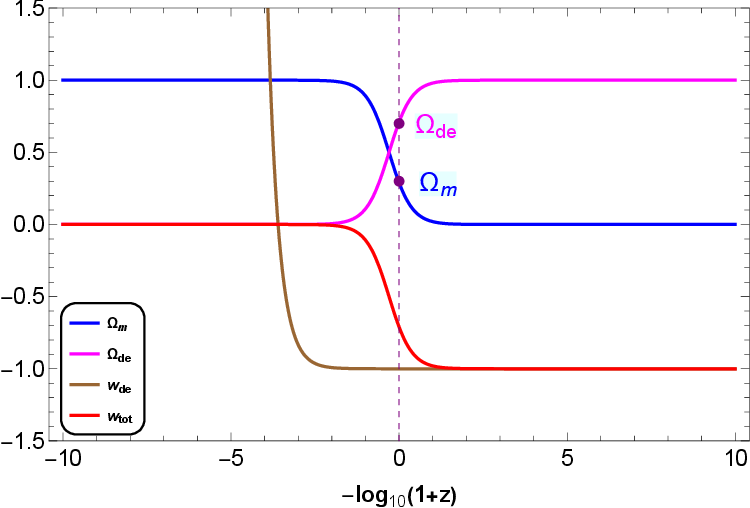}}   
    \hspace{10px}
    \mbox{\includegraphics[scale=0.53]{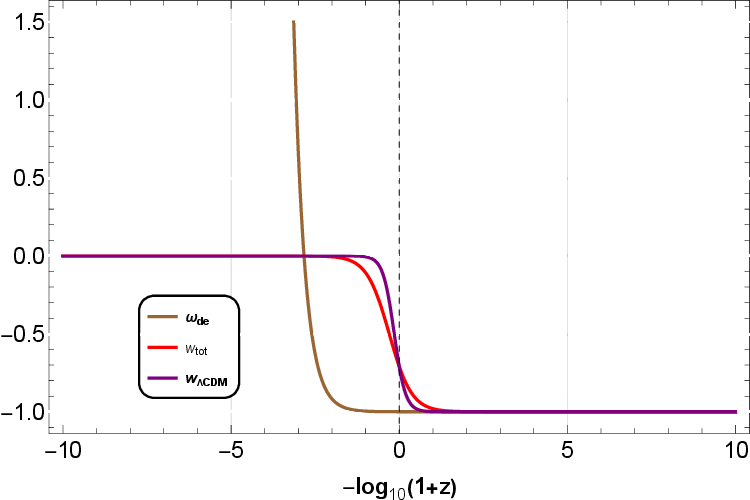}}
    \hspace{10px}
    \mbox{\includegraphics[scale=0.53]{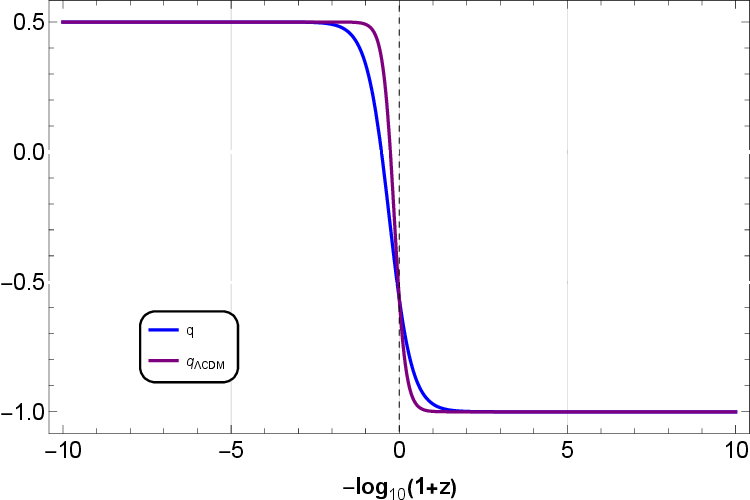}}
    \caption{The evolution of the density parameters (shown in the \textbf{Upper panel}), the EoS parameters (illustrated in the \textbf{Middle panel}), and the deceleration parameter (depicted in the \textbf{Lower panel}) are presented for \textbf{Model-III}. The initial conditions are set as $x = 10^{-3}$, $y = 10^{-6}$, and $\alpha=0.001$. The vertical dashed line represents the present time.}
   \label{Phase Portrait for the dynamical system of Model II}
\end{figure}

Figure 4 (Upper and Middle panels) provides an illustration of the evolutionary history of the density and equation of state (EoS) parameters as functions of redshift $ z $. The initial conditions are calibrated to correspond to present-day values (at $ z = 0 $). In the Upper panel, the transition of the universe from a matter-dominated phase to an accelerated expansion era is depicted. The current density parameters are approximately $ \Omega_{\mathrm{m}} \approx 0.3 $ for matter and $ \Omega_{\mathrm{de}} \approx 0.7 $ for dark energy. The Middle panel shows the evolution of the total EoS parameter, which starts in a matter-dominated era with $ \displaystyle w_{\mathrm{tot}} = 0 $ and evolves towards the dark energy sector, where $ \displaystyle w_{\mathrm{tot}} \approx -1 $. The dark energy EoS parameter approaches $ -1 $ in the later stages of evolution, with the present value of $ \displaystyle w_{\mathrm{de}} = -1 $ aligning with the current observational constraint of $ \displaystyle w_{\mathrm{de}} = -1.028 \pm 0.032 $ \cite{Planck:2018fzr}. In the Lower panel, the deceleration parameter transitions from a decelerating phase to an accelerating phase, with the transition occurring at $ z = 0.64 $ and the current value of the deceleration parameter being $ q_{0} = -0.56 $ \cite{Camarena:2019moy}.

\subsection{MODEL III : $ f(Q)= \kappa Q \ln \left( \frac{Q}{Q_{0}} \right)   $}
In this subsection we consider 
\begin{equation}
F(Q)=\kappa Q \ln \left( \frac{Q}{Q_{0}} \right)- Q ,
\end{equation}
The given $ f(Q) $ model introduces a logarithmic correction to the standard symmetric teleparallel equivalent of General Relativity (STEGR), motivated by quantum gravity effects, renormalization group running, and entropic gravity considerations. The term $ \kappa Q \ln(Q/Q_{0}) $ represents a scale-dependent modification that becomes significant at large cosmic scales while preserving GR at small scales. Such logarithmic terms commonly appear in quantum corrections to gravitational actions and can provide a dynamical explanation for cosmic acceleration without requiring a cosmological constant. The model naturally recovers GR in the limit $ \kappa \to 0 $ and modifies both background evolution and perturbations, making it a viable candidate for addressing dark energy and late-time cosmic acceleration.

As
\begin{center}
$ QF_{QQ}= \kappa $,
\end{center}
The system (24)--(27) becomes
\begin{align}
x'&=\frac{3(1-x-y)(y+2x)}{4 \kappa-y+2}, \\
y'&=\frac{12 \kappa(1-x-y)}{4 \kappa-y+2}, \\
\xi' &= -\xi(\xi+2) + \frac{3(1-x-y)}{2-y} + \frac{3(1-x-y)\xi}{4 \kappa-y+2}.
\end{align}
The given system exhibits singularities only at $ y=2 $.
The corresponding Eos and deceleration parameters reduces to,  
\begin{align}
\displaystyle w_{\mathrm{de}} &= -1 + \frac{(4 \kappa-y)(x+y-1)}{(x+y)(4 \kappa-y+2)}, \\
\displaystyle w_{\mathrm{tot}} &= -1 - \frac{2(x+y-1)}{4 \kappa-y+2}, \\
q &= -1 - \frac{3(x+y-1)}{4 \kappa-y+2}.
\end{align}

Four critical points have been identified and are presented in Table V, which details their associated cosmological characteristics. Table VI provides the eigenvalues of the Jacobian matrix for both the background and perturbation levels.

\begin{table*}
\caption{Critical points (CP), matter, dark energy, perturbation and EoS parameters}
\begin{tabular}{ ||p{1.5cm}||p{2.6cm}||p{0.8cm}||p{0.8cm}||p{0.8cm}||p{2cm}||p{0.8cm}||p{0.8cm}|| }
 \hline
 \textbf{CP} & $\mathbf{(x_{c}, y_{c}, \xi_{c})}$ & $\mathbf{\Omega_{\mathrm{m}}}$ & $\mathbf{\Omega_{\mathrm{de}}}$ & $ \mathbf {\xi} $ & $ \mathbf{\displaystyle w_{\mathrm{de}}} $ & $\mathbf{ \displaystyle w_{\mathrm{tot}}} $ & $ \mathbf{q }$ \\
 \hline
 $ C_{1} $ & $ (x,1-x,0) $ & $ 0 $ & $ 1 $ & $ 0 $ & $ -1+\frac{8\kappa}{1+x+8\kappa} $ & $ -1 $ & $ -1 $  \\
 \hline
 $ C_{2} $ & $ (x,1-x,-2) $ & $ 0 $ & $ 1 $ & $ -2 $ & $ -1+\frac{8\kappa}{1+x+8\kappa} $ & $ -1 $ & $ -1 $  \\
 \hline
  $ C_{3} $ & $ (x,-2x,-\frac{3}{2}) $ & $ 1+x $ & $ -x $ & $ -\frac{3}{2} $ & $ 0 $ & $ 0 $ & $ \frac{1}{2} $ \\
 \hline
  $ C_{4} $ & $ (x,-2x,1) $ & $ 1+x $ & $ -x $ & $ 1 $ & $ 0 $ & $ 0 $ & $ \frac{1}{2} $ \\
 \hline
\end{tabular}
\end{table*} 

\begin{table}
\caption{Eigen values and Stability conditions}
\begin{tabular}{ ||p{1.5cm}||p{2.4cm}||p{3.2cm}|| }
 \hline
 \textbf{CP} & \textbf{Eigen-Values} & \textbf{Stability Condition}  \\
 \hline
 $ C_{1} $ & $ (0,-3,-2) $ & Stable  \\
 \hline
 $ C_{2} $ & $ (0,-3,2) $ & Saddle  \\
 \hline
  $ C_{3} $ & $ (0,\frac{5}{2},3) $ & Unstable \\
 \hline
  $ C_{4} $ & $ (0,-\frac{5}{2},3) $ & Saddle \\
 \hline
\end{tabular}
\end{table} 

 \vspace{5mm} \textbf{* Critical Point $C_{1}=(x,1-x,0)$ :}
The critical point $ B_{4} $ describes a cosmological model characterized by a universe where dark energy completely dominates, with $\Omega_{\mathrm{de}} = 1$ and $\Omega_{\mathrm{m}} = 0$. This implies that the energy density is solely attributed to dark energy, and matter does not contribute significantly to the dynamics of the universe at this critical point with the constant matter perturbation. For the dark energy equation of state parameter, we have:
\begin{center}$ \displaystyle w_{\mathrm{de}} = -1 + \frac{8\kappa}{1 + x + 4\kappa},$ \end{center}
where $ \kappa $ is a constant. This expression suggests that the effective equation of state for dark energy deviates from the cosmological constant value of $-1 $ depending on the values of $ x $ and $ \kappa $. The total equation of state parameter is $\displaystyle w_{\mathrm{tot}} = -1 $, which aligns with a cosmological constant scenario where dark energy is the only significant component contributing to the universe's expansion. The deceleration parameter is $ q = -1 $, indicating an exponential expansion typical of a de Sitter-like universe, where the expansion rate is constant and accelerating. The eigenvalues associated with this critical point are $ (0, -3, -2) $. The presence of a zero eigenvalue signifies that the stability analysis requires further examination through the center manifold theory. The negative eigenvalues suggest stability in the corresponding directions.

The eigenvalue analysis suggests partial stability, with further investigation required to fully understand the critical point's stability in the context of perturbations and dynamical behavior.
In this framework, $x$ functions as the central variable, while $(y, \xi)$ serve as the stable variables. The associated matrices $A$ and $B$ are defined as $A = 0$ and $B = \begin{pmatrix} -3 & 0 \\ 0 & -2 \end{pmatrix}$, respectively. The structure of the center manifold takes the form $y = h_{5}(x)$ and $\xi = h_{6}(x)$, with the approximation $N$ comprising two components.
\begin{flushleft}
$ N_{5}(h_{5}(x))=h_{5}'(x)\frac{3(1-x-h_{5}(x))(h_{5}(x)+2x)}{4 \kappa- h_{5}(x)+2}- \frac{12 \kappa(1-x-h_{5}(x))}{4 \kappa-h_{5}(x)+2}$,
\end{flushleft}
\begin{flushleft}
$ N_{6}(h_{6}(x))=h_{6}'(x) \frac{3(1-x-h_{5}(x))(y+2x)}{4 \kappa-h_{5}(x)+2}+ h_{6}(x)(h_{6}(x)+2) - \frac{3(1-x-h_{5}(x))}{2-h_{5}(x)} - \frac{3(1-x-h_{5}(x))h_{6}(x)}{4 \kappa-h_{5}(x)+2}$.
\end{flushleft}
 \vspace{5mm} \textbf{For zeroth approximation:}\\ $ N_{5}(h_{5}(x))=-\frac{12 \kappa(1-x)}{4 \kappa +2}+\mathcal{O}(x^{2}) $ and\\ $ N_{6}(h_{6}(x))=-\frac{3}{2}(1-x)+\mathcal{O}(x^{2}) $.
\\Therefore the reduced equation gives us
\begin{flushleft}
$ x'=\frac{3 \kappa (50x-15)}{2(11-3x)}+O(x^{2}) $
\end{flushleft}
This analysis reveals a negative linear aspect applicable for $ x \in (-\infty, 0] $. As a result, the system of equations (45)–(47) exhibits asymptotic stability at the equilibrium point, in accordance with the center manifold theory.

Overall, the critical point $ C_{1} $ represents a scenario where dark energy completely governs the universe's dynamics, with specific implications for the equation of state and the expansion history.
 
\vspace{5mm} \textbf{* Critical Point $ C_{2}=(1-y,y,-2) $: }  
The critical point $ C_{2}$ represents a state in the cosmological model where the matter density parameter $\Omega_{\mathrm{m}}$ is zero, indicating the absence of matter, and the dark energy density parameter $\Omega_{\mathrm{de}} = 1$ is fully dominant. This configuration suggests a universe driven entirely by dark energy, with no contribution from matter. The equation of state parameter for dark energy, $ \displaystyle w_{\mathrm{de}} $, is given by $ \displaystyle w_{\mathrm{de}} = -1 + \frac{8\kappa}{1 + x + 4\kappa} $. This indicates a dynamic dark energy model, where $ \displaystyle w_{\mathrm{de}} $ depends on the parameters $ x $ and $ \kappa $. The total equation of state parameter $ \displaystyle w_{\mathrm{tot}} = -1 $ implies a universe that is undergoing accelerated expansion, consistent with a cosmological constant-like behavior with decay in the matter perturbation ($ \xi = -2 $). The deceleration parameter $ q = -1 $ further confirms this, indicating that the universe is in a phase of exponential expansion, characteristic of a de Sitter universe. The eigenvalues associated with this critical point are $ (0, -3, 2) $, suggesting a saddle point behavior. 

This analysis of the critical point $ C_{2} $ highlights the interplay between the dynamic equation of state for dark energy and the stability characteristics, providing insights into the late-time cosmological evolution under this modified gravity theory.

\vspace{5mm} \textbf{* Critical Point $ C_{3}=(x,-2x,-\frac{3}{2}) $: }
The critical point $C_{3} $) describes a cosmological scenario where the matter density parameter is given by $\Omega_{\mathrm{m}} = 1 + x$, and the dark energy density parameter is $\Omega_{\mathrm{de}} = -x$. This configuration suggests a universe where matter dominates, as $\Omega_{\mathrm{m}}$ is positive and greater than 1, while dark energy has a negative density, indicating an unusual or exotic form of energy. Hence, therefore $ x \in [-1,0] $. The equation of state parameter for dark energy, $ \displaystyle w_{\mathrm{de}} = 0 $, indicates that the dark energy behaves like pressureless dust, which is atypical for dark energy but can occur in specific modified gravity scenarios. The total equation of state parameter $ \displaystyle w_{\mathrm{tot}} = 0 $ suggests that the overall universe behaves like a matter-dominated universe, leading to a standard decelerating expansion phase. The deceleration parameter $q = \frac{1}{2}$ supports this interpretation, as it indicates a decelerating universe, characteristic of a matter-dominated phase in the standard cosmological model.

The eigenvalues associated with this critical point are $(0, \frac{5}{2}, 3) $. The presence of positive eigenvalues implies that this critical point is unstable, with any perturbations around this point likely leading to a departure from this state. The perturbation $\xi = -\frac{3}{2}$ further suggests that deviations from this critical point will grow, confirming the instability of this cosmological configuration. In summary, the critical point $ C_{3} = (x, -2x, -\frac{3}{2}) $ describes a matter-dominated universe with an unstable equilibrium, where dark energy behaves unusually as pressureless dust. This instability points to the critical point being a transient state, with the universe likely evolving away from this configuration over time.

\vspace{5mm} \textbf{* Critical Point $ C_{4}=(0,0,1) $: } 
The critical point $ C_{4} $ represents a scenario where the matter density parameter is given by $\Omega_{\mathrm{m}} = 1 + x$ and the dark energy density parameter by $\Omega_{\mathrm{de}} = -x$, with $x \in [-1,0]$. At this point, the equation of state parameters for dark energy ($\displaystyle w_{\mathrm{de}}$) and the total equation of state ($\displaystyle w_{\mathrm{tot}}$) are both zero, indicating a non-evolving dark energy component and a universe dominated equally by matter and dark energy. The deceleration parameter $q = 1/2$ suggests that the universe is in a decelerated expansion phase. The eigenvalues associated with this critical point are $(0, -5/2, 3)$, indicating that the system has a saddle behavior with growth in the matter perturbation. Given these characteristics, the critical point $ C_{4} $ is a saddle point in the dynamical system, with trajectories attracted in some directions and repelled in others. 

\begin{figure}
    \centering
    \includegraphics[scale=0.53]{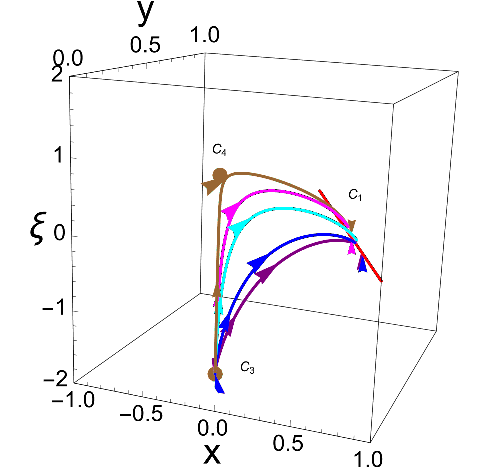}
    \caption{3D phase  portrait for \textbf{Model-III}.}
\end{figure}

The analysis reveals the presence of two matter-dominated critical points ($C_{3}$ and $C_{4}$) and two dark energy-dominated critical points ($C_{1}$ and $C_{2}$) within the framework of the logarithmic form of $ f(Q) $ gravity. The matter-dominated critical points, $C_{3}$ and $C_{4}$, exhibit inherent instability. In particular, critical point $C_{3}$, functioning as a saddle point, signifies a well-defined growth rate in matter perturbations. In contrast, critical point $C_{4}$, identified as an unstable node, represents the decay of matter perturbations, with accelerated expansion observed only at the background level. On the other hand, critical point $C_{1}$ displays consistent accelerated expansion at both the background and perturbation levels, establishing it as a stable configuration.

Figure 5 presents the phase portrait in three-dimensional space, illustrating the trajectory's evolution as it progresses from matter-dominated to dark-energy-dominated critical points. The diagram clearly delineates the sequential transition of the trajectory, commencing at the unstable node associated with critical point $C_{3}$, moving through the saddle point instability at $C_{4}$, and ultimately converging to the stable node represented by critical point $C_{1}$.

\begin{figure} 
   \centering 
   \mbox{\includegraphics[scale=0.53]{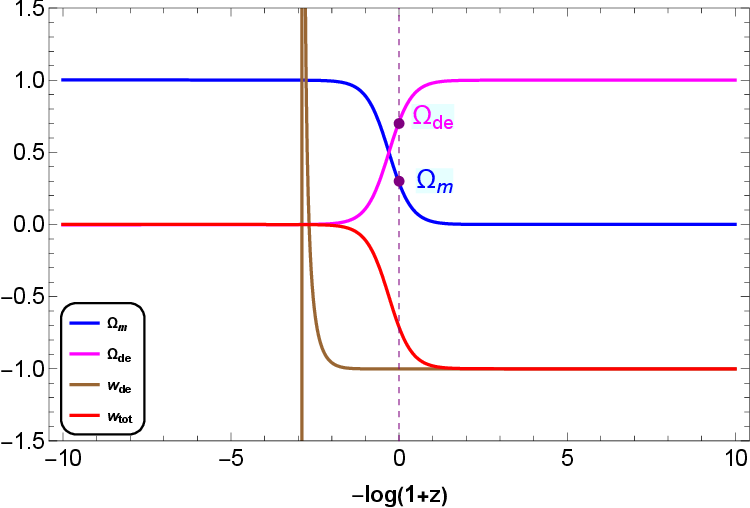}}   
    \hspace{10px}
    \mbox{\includegraphics[scale=0.53]{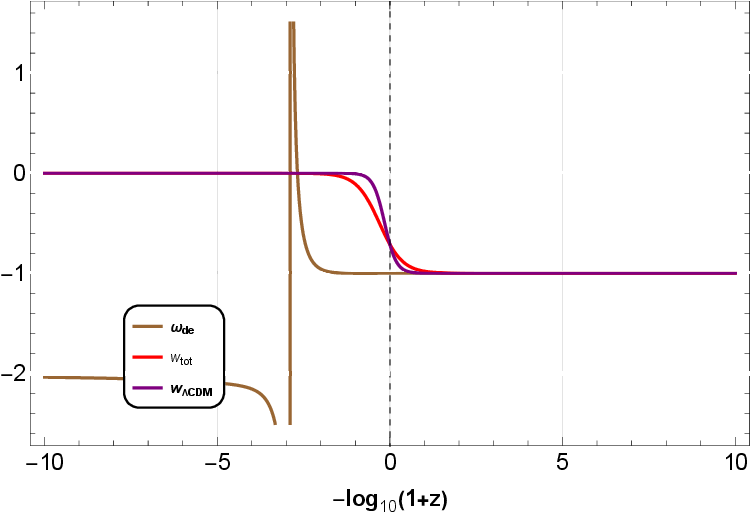}}
    \hspace{10px}
    \mbox{\includegraphics[scale=0.53]{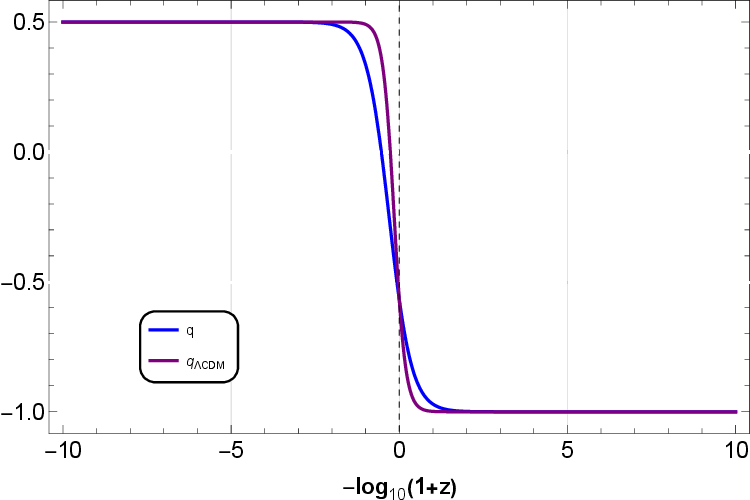}}
    \caption{ The evolution of the density parameters (shown in the \textbf{Upper panel}), the EoS parameters (illustrated in the \textbf{Middle panel}), and the deceleration parameter (depicted in the \textbf{Lower panel}) are presented for \textbf{Model-III}. The initial conditions are set as $x = 10^{-3}$, $y = 10^{-6}$, and $\kappa = 0.0001$. The vertical dashed line represents the present time.}
   \label{Phase Portrait for the dynamical system of Model III}
\end{figure}

Figure 6 (Upper and Middle panels) presents the evolutionary trajectories of the density and equation of state (EoS) parameters as functions of redshift $ z $. The initial conditions are adjusted to match the current values at $ z = 0 $. The \textbf{Upper panel} illustrates the universe's progression from a matter-dominated phase to a phase of accelerated expansion. At present, the density parameters are approximately $\Omega_{\mathrm{m}} \approx 0.3$ for matter and $\Omega_{\mathrm{de}} \approx 0.7$ for dark energy. The \textbf{Middle panel} depicts the evolution of the total equation of state (EoS) parameter, which begins in the matter-dominated regime with $\displaystyle w_{\mathrm{tot}} = 0$ and gradually transitions towards the dark energy regime, where $\displaystyle w_{\mathrm{tot}}$ approaches $-1$. Simultaneously, the EoS parameter for dark energy converges towards $-1$ in the later stages of evolution, aligning with the current observational constraint of $\displaystyle w_{\mathrm{de}} = -1.028 \pm 0.032$ \cite{Planck:2018fzr}. The \textbf{Lower panel} shows the deceleration parameter's shift from a decelerating phase to an accelerating phase, with the transition occurring at $z = 0.60$. The current value of the deceleration parameter is $q_{0} = -0.57$ \cite{Camarena:2019moy}.

\section{Discussion and Conclusion}

The $ f(Q) $ gravity framework provides a rich and versatile approach to modifying GR. By exploring various forms of $ f(Q) $, one can address different cosmological phenomena, from the early universe to late-time acceleration. Further research into the perturbation and observational aspects of $ f(Q) $ gravity will help in understanding its viability as an alternative to GR.

Dynamical system analysis serves as a valuable tool for exploring the qualitative behavior of the universe. This method involves addressing non-linear differential equations through the framework of dynamical variables, thereby characterizing the universe's evolution via the critical points of autonomous systems. In this study, we employed dynamical system analysis within the context of $ f(Q) $ gravity, examining both background and perturbation levels. Specifically, we formulated the general autonomous dynamical systems (equations (19)–(21)) within the symmetric teleparallel framework, focusing on $ f(Q) $ gravity. Here, the dynamical variables $ x $ and $ y $ represent the background evolution of the universe, while the variable $ \xi $ captures the perturbative aspects, including the growth and decay of matter perturbations. The autonomous systems we defined incorporate the functional form of $ f(Q) $, leading to the proposal of three distinct models based on different forms of $f(Q)$.

In Model-I, we examined a logarithmic form of $ f(Q) $ as presented in equation (22). This approach identified four critical points, which describe the matter-dominated and dark energy-dominated phases of the Universe at both the background and perturbation levels. Critical points $A_{1}$ and $A_{2}$ correspond to the dark energy-dominated era, with $A_{2}$ exhibiting accelerated expansion at the background level and decay in matter perturbations, while $A_{1}$ displays accelerated expansion and stable node behavior at both levels. Conversely, critical points $A_{3}$ and $A_{4}$ are associated with the matter-dominated era, where $A_{3}$ indicates a growth rate in matter perturbations, while $A_{4}$ signals their decay.
In Model–II, we considered a polynomial form of $ f(Q) $, presented in equation (30), which also produced four critical points. The qualitative behavior of these critical points is similar to that observed in Model–I, despite the different functional forms of $ f(Q) $. Here, critical points $B_{1}$ and $B_{2}$ describe the dark energy-dominated phase, with $B_{1}$ uniquely illustrating late-time cosmic acceleration at both levels, whereas $B_{3}$ and $B_{4}$ define the matter-dominated phase.
Model–III revisits the logarithmic form of $ f(Q) $, as given in equation (38), and similarly identifies four critical points. The behavior of these critical points mirrors that of the previous models, maintaining consistency across different functional forms of $ f(Q) $. The critical points $ C_{1} $ and $ C_{2} $ characterize the dark energy-dominated phase, with $ C_{1} $ distinctively demonstrating late-time cosmic acceleration at both the background and perturbation levels, while $ C_{3} $ and $ C_{4} $ are associated with the matter-dominated phase.

The qualitative dynamics of this model align with those of Models I, II and III, both at the background and perturbation levels. Notably, cosmological perturbation analyses have been conducted to assess the stability of cosmological models in $ f(Q) $ gravity, as discussed in Refs. \cite{Coley:2023dbg, Coley:2023kyg}. These investigations focus on a class of Einstein teleparallel geometries characterized by a four-dimensional Lie algebra of affine connections, with explicit forms of $ f(Q) $ derived for various parameter values. Our study considers three such forms of $ f(Q) $ to demonstrate the Universe's late-time cosmic acceleration through dynamical system analysis.

The cosmological evolution of the Universe has been assessed by analyzing the density parameters for matter and dark energy, the equation of state (EoS) parameters, and the deceleration parameters. Across all models, the deceleration parameter consistently indicates a transition from early-time deceleration to late-time acceleration, with the transition occurring at $ z = 0.59 $, $ z = 0.64 $, and $ z = 0.60 $, respectively. The corresponding present-day values of the deceleration parameter are $ q_{0} = -0.57 $, $ q_{0} = -0.56 $, and $ q_{0} = -0.57 $. All three models yield the same present-day value for the dark energy EoS parameter, $ \displaystyle w_{\mathrm{de}} = -1 $. Moreover, the density parameters for matter and dark energy are determined to be $ \Omega_{\mathrm{m}} \approx 0.3 $ and $ \Omega_{\mathrm{de}} \approx 0.7 $, respectively, consistent with current cosmological observations. Phase space trajectories have been constructed in three-dimensional space for each model, illustrating the transition from an unstable, matter-dominated phase to a stable, dark energy-dominated phase. This study concludes that dynamical stability analysis is a valuable tool for extensively investigating the cosmological behavior of the Universe. In $ f(Q) $ gravity, the identification of such a critical point is significant as it demonstrates the possibility of achieving a stable accelerated expansion driven by the modified gravity framework, without the need for additional exotic matter components.

\section*{Acknowledgement}
We sincerely appreciate the constructive and valuable suggestions and comments provided by the anonymous referees, which have significantly contributed to improving the quality of the manuscript.

\section*{Acknowledgement}
The work of KB was supported by the JSPS KAKENHI Grant Numbers 21K03547, 23KF0008, 24KF0100.

\end{document}